# An End-to-End Approach for Korean Wakeword Systems with Speaker Authentication

Geonwoo Seo
Dongguk University

## ABSTRACT

Wakeword detection plays a critical role in enabling AI assistants to listen to user voices and interact effectively. However, for languages other than English, there is a significant lack of pre-trained wakeword models. Additionally, systems that merely determine the presence of a wakeword can pose serious privacy concerns. In this paper, we propose an end-to-end approach that trains wakewords for Non-English languages, particulary Korean, and uses this to develop a Voice Authentication model to protect user privacy. Our implementation employs an open-source platform OpenWakeWord, which performs wakeword detection using an FCN (Fully-Connected Network) architecture. Once a wakeword is detected, our custom-developed code calculates cosine similarity for robust user authentication. Experimental results demonstrate the effectiveness of our approach, achieving a 16.79% and a 6.6% Equal Error Rate (EER) each in the Wakeword Detection and the Voice Authentication. These findings highlight the model's potential in providing secure and accurate wakeword detection and authentication for Korean users.

## INTRODUCTION

The evolution of voice-based assistive technologies has reshaped the way users interact with artificial intelligence models. Modern AI-driven voice assistants—often leveraging advanced generative AI—can understand complex natural language commands and perform diverse tasks. A critical enabler of these systems is wakeword detection, which ensures that the AI assistant only listens when a specific keyword is uttered. Yet, most public wakeword detection models still focus heavily on English, leaving significant gaps in support for other languages such as Korean. This imbalance not only hinders accessibility for non-English speakers but also limits the applicability of voice-based AI in global contexts[1].

Moreover, the always-listening nature of wakeword-based solutions raises concerns about unauthorized access, as anyone who knows or accidentally speaks the wakeword may activate the system. As generative AI capabilities expand and become increasingly proactive, a robust authentication layer becomes essential. Our work addresses these issues by proposing an end-to-end methodology for Korean wakeword detection coupled with a speaker authentication step. The overall process involves four key units working in sequence:

1. Pre-processing converts raw audio into mel-spectrograms.
2. A shared feature extraction backbone transforms these mel-spectrograms into compact speech embeddings.
3. A wakeword classification module identifies whether the uttered phrase matches the designated wakeword.
4. A speaker authentication module verifies the identity of the speaker and grants access only to authorized individuals.

Through this two-step verification—wakeword detection and speaker authentication—we aim to provide a fast, accurate, and privacy-focused voice interface for Korean-language users. By integrating a Korean wakeword, employing lightweight models, and utilizing data augmentation (e.g., TTS, background noise, and reverberation), our system demonstrates robust performance in real-world environments. Additionally, leveraging the CUDA platform enables efficient on-device inference, making this approach suitable for resource-constrained devices as well as mainstream AI assistant deployments.

# SYSTEM DESIGN AND ARCHITECTURE

In this work, we extend the open-source OpenWakeWord platform to detect and authenticate a Korean wakeword for enhanced privacy and accessibility[2]. Specifically, we adapt the system to recognize the wakeword "하나", chosen for its relatively simple phonetic structure and partial resemblance to certain English phonemes. This choice not only streamlines model adaptation but also enables text-to-speech (TTS) data augmentation—we can use TTS-generated "hana" audio samples to diversify our training set without requiring an entirely separate large-scale dataset. Our method comprises a structured processing pipeline that sequentially detects the wakeword and verifies the speaker's identity, ensuring both efficiency and security.

## 2.1 Processing Pipeline

**Pre-processing Unit**
We continuously monitor the incoming audio at a 16 kHz sampling rate. Once the system deems the audio ready for processing, it segments the stream into manageable chunks. These chunks are then converted into mel-spectrograms, preserving essential spectral features crucial for both detecting the wakeword and performing subsequent speaker authentication.

**Shared Feature Extraction Backbone**
The mel-spectrograms pass into a lightweight convolutional neural network (adapted from a TFHub module), where they are converted into 96-dimensional embeddings. This backbone is trained on large-scale audio data for generalization and then frozen to reduce computational overhead during inference. Incorporating data augmentation methods—such as reverberation (using MIT's Room Impulse Responses) and various background noises—further improves robustness to real-world conditions.

**Wakeword Classification**
The 96-dimensional embeddings flow into a fully-connected classifier that identifies whether the utterance contains the target wakeword "하나". We employ gradient accumulation to simulate large batch sizes during training without exceeding memory constraints. The classifier outputs a wakeword detection score, indicating the probability of the target keyword's presence.

**Speaker Authentication**
Once a wakeword is detected, the system proceeds to verify the speaker's identity. Depending on the configuration, the system follows one of two approaches. In Approach A (Parallel Pipeline), the 96-dimensional speaker embedding obtained during wakeword detection is directly used for verification. This method allows for a faster authentication process since it does not require additional embedding computation. The flow of this approach is illustrated in Figure 1.

In contrast, Approach B (Post-classification Trigger) generates a higher-dimensional (256-dimensional) speaker embedding from a local audio cache after wakeword detection. This additional step enables a more refined verification process before granting access, potentially improving accuracy at the cost of a slight increase in latency. The flow of this approach is shown in Figure 2.

In both approaches, the system compares the speaker's embedding to a known authorized-user model using a cosine similarity metric. If the similarity exceeds a predefined threshold, the AI assistant is fully activated.

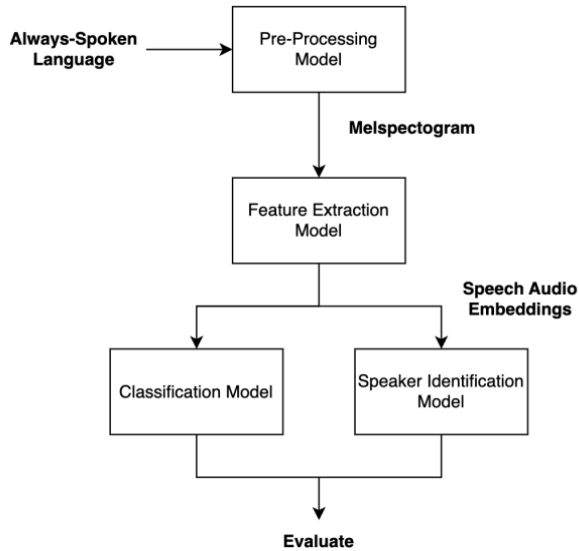

Figure 2 Flow chart based on Approach A

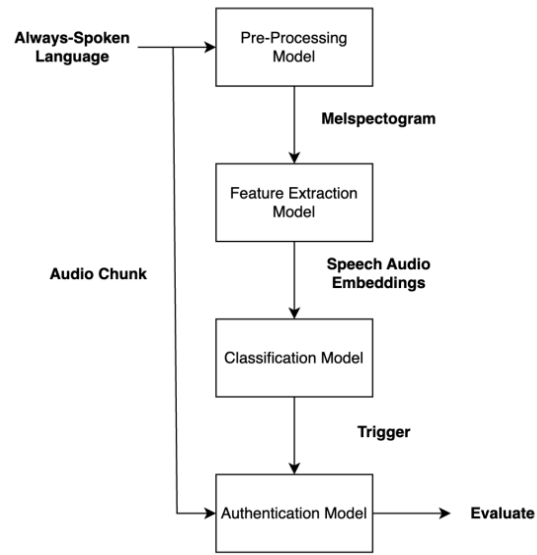

Figure 1 Flow chart based on Approach B

## 2.2 Korean Wakeword Selection

By incorporating a Korean-specific wakeword, leveraging TTS-based data augmentation for diverse training samples, and applying lightweight yet robust modeling techniques, our pipeline addresses both linguistic inclusivity and privacy concerns. Additionally, integrating the CUDA platform enables parallel computations and reduces latency, making our approach suitable for real-time deployment on resource-constrained devices. Since OpenWakeWord primarily supports English wakewords, we first adapted the pipeline for the Korean wakeword "하나". This choice was motivated by the relatively simple phonetic structure, as well as its partial resemblance to certain English phonemes. This allows us to utilize TTS-generated "hana" audio samples for data augmentation.

## MODULE IMPLEMENTATION DETAILS

### 3.1 Integrate OpenWakeWord platform

OpenWakeWord processes incoming audio through a structured three-stage pipeline, with each stage operating within separate threads. Each stage incrementally transforms input data and forwards the processed output to the next stage. This modular approach ensures efficient wake word detection while maintaining low latency. Below we will explain operating details and parameters of each stages.

**Pre-processing Unit (mels_proc)**
The first stage of processing involves transforming raw audio input into mel spectrogram representations. This stage continuously receives 16 kHz audio samples from each client, accumulating them until a 1,760 sample being stored. The accumulated audio is then fed into a TensorFlow Lite (TFLite) model (melspectrogram.tflite), which computes the mel spectrogram frames.

The model performs a Short-Time Fourier Transform (STFT) followed by mel-filtering, outputting a spectrogram with dimensions [batch_size, 1, window_count, 32]. The window size for STFT is 25 milliseconds, corresponding to 400 samples, ensuring overlapping frames for smooth spectral transitions. The mel spectrogram consists of 32 mel bins, capturing frequency components in the range of approximately 60 Hz to 3800 Hz.

To normalize the magnitudes for compatibility with the embedding model, a scaling transformation is applied: (mels / 10) + 2. The computed mel spectrogram frames are appended to each client's ring buffer (client.mels), and the variable client.new_mels is updated to reflect the availability of new frames.

**Shared Feature Extraction Backbone (embeddings_proc)**

The second stage extracts embeddings from the mel spectrogram frames. Once a client's buffer accumulates at least EMB_FEATURES of new mel frames (client.new_mels ≥ EMB_FEATURES), a batch is formed and processed by a second TFLite model (embedding_model.tflite). This model generates embeddings, which are 96-dimensional feature vectors.

The input to the embedding model has the shape [batch_size, EMB_FEATURES, 32, 1], where EMB_FEATURES represents the number of mel frames required per embedding extraction. The output consists of embeddings with dimensions [batch_size, 1, embedding_windows, 96], where each embedding has 96 features. These extracted embeddings are stored in the wake word buffer, and the client.new_embeddings counter is updated accordingly.

To maintain efficiency, old embeddings are overwritten by new ones in a ring buffer system. This process ensures continuous adaptation to incoming audio while keeping memory usage stable. The mel-to-embedding transformation is performed only once and shared across multiple wake words to minimize redundant computation.

**Wakeword Classification (ww_proc)**

The final stage evaluates the presence of wake words by processing embeddings from the client's buffer. For each registered wake word (ww_model_key), a fixed window of embeddings (ww_windows) is extracted and input into a TFLite classification model. This model takes an input of shape [1, ww_windows, 96] and outputs a probability score indicating the likelihood of a wake word being detected.

When client_data.new_embeddings reaches or exceeds the required number of embeddings for classification (client_data.new_embeddings ≥ ww_windows), a batch is processed by the wake word model. The model's output probability is then compared against a predefined threshold (client_data.wake_threshold). If the probability exceeds this threshold, an activation counter (client_data.activations) is incremented. To confirm wakeword detection, the system requires a certain number of consecutive positive detections, defined by client_data.trigger_level. Once this level is reached, the system triggers a speaker identification model to verify the speaker's identity.

## 3.2 Speaker Authentication Strategy

In addressing the privacy gap where any user uttering the wakeword could inadvertently trigger the assistant and gain access to private information, two distinct speaker authentication strategies are proposed, each characterized by specific parameters such as wake thresholds, authentication thresholds, and temporal constraints to ensure both accuracy and responsiveness.

**Approach A (Parallel Pipeline)**

This strategy leverages the existing 96-dimensional embeddings produced during wakeword detection. As these embeddings are fed into the classifier, a concurrent verification pathway utilizes the same embedding to ascertain the speaker's identity. The speaker verification runs in parallel by comparing the same low-dimensional embedding against a stored reference using a cosine similarity measure. The authentication is considered successful only if the computed similarity surpasses both authentication threshold (auth_threshold) and the wakeword threshold (wake_threshold). This dual-threshold gating mechanism ensures fast and efficient verification, but the limited dimensionality of the embedding may pose challenges in speaker discrimination under certain conditions.

**Approach B (Post-classification Trigger)**

In this approach, once the system detects a wakeword with a probability exceeding the wake_threshold, it temporarily halts further processing in a cooldown period (20 frames) to avoid repeated triggers. During this interval, the system retrieves a 4,000 recent audio frames (0.25 seconds) from a local cache. This audio chunk is then processed by an encoder to produce a more expressive 256-dimensional speaker embedding, which is better suited for capturing the intricacies of the speaker's voice. The similarity between this newly computed embedding and a stored reference embedding for the authorized speaker is then calculated using cosine similarity. Only if the similarity exceeds auth_threshold does the system confirm the speaker's identity and proceed to fully activate the assistant. The use of a higher-dimensional embedding at this stage reduces false activations by non-authorized speakers while maintaining user convenience.

### 3.3 Implementation Details

All experiments were conducted within the OpenWakeWord framework, where we adapted various models and threshold settings to improve wake word detection accuracy for Korean, specifically for the wakeword "하나". Our primary objective was to balance high detection accuracy with real-time responsiveness, ensuring a smooth user experience in AI assistant applications.

**Parallel Computing and CUDA Integration**

To enhance inference speed, we integrated the CUDA platform into the pipeline, allowing the speaker verification model to efficiently utilize GPU acceleration. By leveraging parallel computation, we significantly reduced latency, enabling near-instantaneous wake word detection and verification on compatible hardware.

**Reverberation and Background Noise Augmentation**

We first introduced reverberation effects using MIT's Room Impulse Responses (RIR) dataset, allowing us to simulate a variety of room acoustics. This helped the model adapt to echoes and reflections, making it more resilient to real-world reverberant environments.

Additionally, we incorporated background noise augmentation by using publicly available noise datasets, which included ambient sounds such as café chatter, traffic noise, and home appliance sounds. These augmentations improved the model's robustness against challenging acoustic conditions.

**Audiomentations-Based Augmentation**

We utilized the audiomentations library to introduce diverse audio distortions that improve robustness.
- SevenBandParametricEQ: Adjusts frequency bands to simulate variations in microphone characteristics.

- TanhDistortion: Introduces controlled distortion to improve robustness against low-quality recordings.
- PitchShift: Randomly alters pitch to make the model resilient to natural voice pitch variations.
- BandStopFilter: Removes specific frequency bands to simulate hardware-related limitations or low-quality audio sources.
- AddColoredNoise: Injects colored noise at varying signal-to-noise ratios (SNRs), simulating diverse background conditions.
- Gain Augmentation: Modifies audio loudness, ensuring wake words can be detected at different volume levels.

Through this end-to-end methodology, we successfully customized OpenWakeWord to detect the Korean wake word "하나" while incorporating a robust speaker authentication mechanism. By utilizing lightweight classification and verification models, our system achieves high detection accuracy while maintaining low-latency inference, a crucial factor for AI assistants in real-time environments.

Additionally, our extensive data augmentation pipeline, incorporating reverberation modeling, background noise addition, and adaptive transformations using audiomentations, ensures strong performance across diverse audio conditions. The integration of CUDA for GPU acceleration further optimizes performance, enabling fast and scalable wake word detection.

By combining efficiency, adaptability, and security, our system is well-suited for real-world deployment in AI assistants, ensuring accurate and privacy-conscious wake word detection in Korean-language applications.

### 3.4 Pseudocode

The pseudocode below clearly describes the parallel processing flow of the wake word pipeline and distinctly branches the two speaker authentication approaches (A and B). The pseudocode implements branching within conditional statements for Approaches A and B, demonstrating that different authentication procedures are executed depending on the user-selected method.

**Algorithm WakeWordPipeline**

**Input:**
   - TFLite models: mel_spectrogram_model, embedding_model, wakeword_model
   - Audio streams from clients
   - Speaker reference embeddings for authentication
   - Thresholds: wake_threshold, auth_threshold, trigger_level, COOLDOWN_FRAMES, etc.
**Output:** Wake word detection and speaker authentication events

  **Parallel Execute:**
   **Thread PreProcessingUnit:**
    while system is running:
     Wait until sufficient new audio samples (≥ MEL_SAMPLES) are collected for any client
     for each client with new_audio_samples ≥ MEL_SAMPLES:
      audio_batch ← extract latest MEL_SAMPLES samples
      mels ← mel_spectrogram_model.infer(audio_batch)
      Normalize mels: mels ← (mels / 10) + 2
      Append mels to client's mel ring buffer
      Update client.new_mels and timestamp
     Signal SharedFeatureExtraction (mels_ready)

**Thread SharedFeatureExtraction:**
  while system is running:
    Wait for mels_ready signal
    for each client with new_mels ≥ EMB_FEATURES:
      mel_batch ← extract EMB_FEATURES frames from client's mel buffer
      embeddings ← embedding_model.infer(mel_batch)
      for each wake word model ww:
        if client uses ww:
          Append embeddings to client's ww-specific embedding buffer
          Update client.new_embeddings and timestamp
    Signal each WakewordClassification thread (embeddings_ready)

**For each wake word model ww_model_key:**
  **Thread WakewordClassification(ww_model_key):**
    Load wakeword_model and corresponding speaker reference (if needed)
    while system is running:
      Wait for embeddings_ready signal for ww_model_key
      for each client with new_embeddings ≥ ww_windows:
        embeddings_window ← extract ww_windows embeddings from client buffer
        probability ← wakeword_model.infer(embeddings_window)

        if client.cooldown_counter > 0:
          Skip processing due to cooldown period
        else if probability ≥ client.wake_threshold:
          client.activations += 1
          if client.activations ≥ client.trigger_level:
            client.activations ← 0
            client.cooldown_counter ← COOLDOWN_FRAMES

            if approach == A:
              sr, n_mels, num_frames ← 16000, 96, 50
              audio_data ← retrieve recent audio for client
              mel_spec ← librosa.feature.melspectrogram(y=audio_data, sr=sr, n_mels=n_mels, hop_length=512)
              test_emb ← average first num_frames of mel_spec along time axis
              similarity ← cosine_similarity(test_emb, reference_embedding_A)
              if similarity ≥ client.auth_threshold:
                Trigger wake word detection event for client
                Log "Auth Success" with similarity score
              else:
                Log "Auth Failed" with similarity score

            else if approach == B:
              audio_chunk ← extract 4000 recent samples from client's audio buffer
              speaker_embedding ← VoiceEncoder.encode(audio_chunk)
              similarity ← cosine_similarity(speaker_embedding, reference_embedding_B)
              if similarity ≥ client.auth_threshold:
                client.is_detected ← True
                Trigger wake word detection event for client
                Log "Auth Success" with similarity score
              else:
                Log "Auth Failed" with similarity score

        else:
          client.activations ← max(0, client.activations - 1)

```
if client.cooldown_counter > 0:
    client.cooldown_counter -= 1
```

Update client.new_embeddings and timestamp as embeddings are consumed

## EXPERIMENTS

### 4.1 Data Description

Table 1 summarizes the dataset collected from seven individual speakers, resulting in a total of 1,287 audio samples. These samples are stored in WAV format and capture each speaker uttering the Korean word "하나". To ensure speaker diversity, we split the dataset into training (n = 919) and testing (n = 368) sets. Each speaker contributed a varying number of utterances—ranging from 30 to 360 in the training portion—with corresponding test data proportionally assigned.

**Table 1** Dataset Composition by Speaker

| Speaker | Train | Test | Total |
|---|---|---|---|
| JJA | 100 | 46 | 146 |
| LDH | 200 | 80 | 280 |
| LJH | 30 | 16 | 46 |
| NHW | 60 | 32 | 92 |
| PJH | 89 | 28 | 117 |
| SGW | 360 | 131 | 491 |
| SHS | 80 | 35 | 115 |
| Total | 919 | 368 | 1287 |

Although all utterances contain the same wakeword "하나", speaker identities differ. We designate SGW as the authorized user and the remaining six participants as unauthorized users. Consequently, these wakeword-containing samples are divided into two categories:

- voice-authp (wakeword positive, authorized user): Utterances from SGW
- voice-authn (wakeword positive, unauthorized user): Utterances from all other speakers

In addition to the human-recorded samples, we generated 2,000 synthetic audio clips using openai-piper, selecting "haanaa" for its close phonetic resemblance to the Korean word "하나".We then organized these samples into two TTS-based categories:

- tts-wwp (wakeword positives): Contains TTS-generated "haanaa" utterances
- tts-wwn (wakeword negatives): Contains TTS-generated utterances but not including "haanaa".

Lastly, to evaluate the system's false activation rates, we included a Korean conversation dataset containing non-wakeword audio segments from AI-Hub[1]. This dataset was segmented into short units, each containing a single sentence, enabling sentence-level detection/authentication analysis. Consequently, the overall false rejection rate (FRR) and false acceptance rate (FAR) could be expressed as percentages based on individual sentence detection results. The overall dataset composition—including TTS and conversational clips—is summarized in Table 2.

---

[1] https://aihub.or.kr/aihubdata/data/view.do?currMenu=115&topMenu=100&aihubDataSe=realm&dataSetSn=109

Table 2 Train and Test Data Composition by Category

| Category | Train | Test |
|---|---|---|
| voice-authp | 360 | 131 |
| voice-authn | – | 796 |
| tts-wwp | 1000 | – |
| tts-wwn | 1000 | – |
| conversation | 1000 | – |

### 4.2 Data Collection Method

All recorded voice data in Tables 1 and 2 were collected using a custom Python script named recorder.py. We bundled this script into an executable using PyInstaller, which allowed participants to run the recorder on various operating systems without installing Python or additional dependencies. The script automatically records 1-second clips at a 16 kHz sampling rate, separated by a 2-second wait period. Participants can easily pause and resume recording by pressing the space bar, with this functionality managed by a background keyboard listener (pynput). This design enables users to control the recording process and capture multiple wakeword utterances under different acoustic conditions without restarting the entire program.

### 4.3 Preprocessing & Data Augmentation

Prior to training, we performed two core preprocessing steps: normalization and voice activity detection (VAD). We used a custom script (normalize.py[2]) based on the PyDub library to standardize audio amplitudes. For VAD, we employed vad.py [3] to detect and remove silent sections in each WAV file. However, since we wished to investigate VAD's impact on final performance, we retained two variants of each dataset: one with VAD-applied clips and one without. This approach allowed us to compare how removing silence affected both wakeword detection and speaker authentication.

After optional VAD trimming, we augmented the audio clips to improve model robustness. We leveraged audiomentations and torch_audiomentations with fixed probabilities, including 75% for background noise addition, 25% for pitch shifting, and 50% for RIR-based reverberation. This augmentation process helps the model generalize more effectively to real-world acoustic variations, such as echo-prone rooms and noisy environments.

### 4.4 Experimental Setup

In this study, all wakeword models were trained using Google Colab's GPU-accelerated environment to support large-scale experiments, while inference and real-time tests were conducted on an NVIDIA Jetson Nano running Ubuntu 20.04. The Jetson Nano, equipped with a CUDA-capable GPU, allowed partial offloading of speaker authentication tasks to the GPU, resulting in faster response times. The software environment was based on a customized Ubuntu 20.04 image, PyTorch 1.13.0 (with CUDA 10.2), and Python 3.8.

---

[2] https://github.com/gws8820/securewakeword-model/blob/main/voice/normalize.py
[3] https://github.com/gws8820/securewakeword-model/blob/main/voice/vad.py

The model architecture consisted of a preprocessing unit with a feature extraction backbone, a wakeword classifier, and a speaker authentication module. The preprocessing unit, feature extraction backbone, and wakeword classifier all operated on the CPU, due to incompatible CUDA version, whereas the speaker authentication process was accelerated on the GPU using CUDA. For wakeword model training, positive samples were prepared using two combinations of datasets. The first combination, **voice_919**, merged voice-authp and voice-authn, and the second combination, **mix_919**, combined voice-authp, voice-authn, and tts-wwp. Negative samples were drawn from the tts-wwn dataset. Additionally, experiments were conducted both with and without applying Voice Activity Detection (VAD) to trim silence, in order to evaluate the impact of silence removal on the training process. A detailed explanation is provided in the Table 3.

Table 3 Comaprison Table of Wakeword models

| Model | Positive Samples | Negative Samples | VAD Applied |
|---|---|---|---|
| voice_919 | voice-authp voice-authn | tts-wwn | No |
| voice_919_vad | voice-authp voice-authn | tts-wwn | Yes |
| mix_919 | voice-authp voice-authn tts-wwp | tts-wwn | No |
| mix_919_vad | voice-authp voice-authn tts-wwp | tts-wwn | Yes |

As we obeserved, for training the speaker authentication model, two approaches were taken. Approach A directly used the 96-dimensional embeddings provided by the shared backbone. In Approach B, 256-dimensional embeddings were generated using three different models—**Resemblyzer**, **x-vector**, and **ECAPA**—to compare speaker verification performance. Table 4 provides a detailed comparison of these speaker authentication models, highlighting their embedding dimensions and architectural characteristics.

Table 4 Comparison Table of Speaker Authentication Models

| Model | Embedding Dimension | Architecture |
|---|---|---|
| Resemblyzer | 256-Dimensional | Fixed-Dimensional Embeddings |
| x-vector | 256-Dimensional | Time-Delay Neural Networks (TDNN) |
| ECAPA | 256-Dimensional | TDNN-based channel attention |

Resemblyzer is a deep learning-based tool designed for voice similarity tasks. It converts raw audio data into fixed-dimensional speaker characteristic embeddings. By effectively extracting features from the speech, Resemblyzer is useful for applications such as speaker identification, voice conversion, and speaker separation.

The x-vector model is a deep learning architecture that employs time-delay neural networks (TDNNs) to learn speaker-specific characteristics along the temporal axis[4]. Trained on large-scale labeled speaker data, the x-vector approach extracts embeddings that are widely used as a standard method in speaker verification, capturing robust speaker representations.

ECAPA (Emphasized Channel Attention, Propagation, and Aggregation) is a recent advancement in speaker verification technology[5]. It enhances feature representation by emphasizing channel attention mechanisms within the network, refining the flow of information. As a result, ECAPA produces strong and discriminative 256-dimensional speaker embeddings, significantly improving speaker verification performance.

## 4.5 Evaluation Metrics

The False Rejection Rate (FRR) is calculated using the formula:

$$\text{FRR} = \frac{FN}{FN + TP}$$

This represents the proportion of actual positive cases that are incorrectly classified as negative by the model. Here, FN (False Negatives) denotes the number of positive instances that the model fails to recognize, while TP (True Positives) denotes the number of positive instances that the model correctly identifies. Because the denominator (FN + TP) covers all actual positive cases, **calculating FRR requires a positive dataset** that contains examples where the event truly occurs. This metric reflects how often the system misses true positive events, such as failing to detect the wakeword when it is present or not recognizing an authorized speaker when they speak.

The False Acceptance Rate (FAR) is given by the formula:

$$\text{FAR} = \frac{FP}{FP + TN}$$

FAR measures the proportion of actual negative cases that the model incorrectly classifies as positive. In this equation, FP (False Positives) is the number of negative instances that are wrongly identified as positive, and TN (True Negatives) is the number of negative instances that the model correctly classifies. Since the denominator (FP + TN) encompasses all actual negative cases, **computing FAR requires a negative dataset** that contains examples where the event of interest is absent. A high FAR would mean that the system often falsely identifies the absence of a wakeword as its presence or mistakenly authenticates an unauthorized speaker.

In addition to FRR and FAR, the Equal Error Rate (EER) is a key metric for evaluating the balance between these two types of errors. EER is defined as the error rate at which the False Acceptance Rate and the False Rejection Rate are equal[3]. To determine the EER, one varies the decision threshold of the model and computes FAR and FRR at each threshold. **The EER is found at the threshold where FAR equals FRR**. A lower EER indicates a more accurate and balanced model performance.

To calculate these metrics correctly, the dataset must consist of specific types of data that represent positive and negative cases. For FRR, a positive dataset is required. In the context of the wakeword detection model, this positive dataset might include audio samples from speakers saying the Korean wakeword "하나".

For instance, using the voice_919 combination, positive cases would come from both authorized (voice-authp) and unauthorized (voice-authn) users uttering "하나" while the mix_919 combination would additionally include TTS-generated "hana" samples (tts-wwp). These samples represent true positive scenarios for the model, where the wakeword is present and should be detected. The FRR is then computed by running the model on these positive samples, counting how often the wakeword is rejected (False Negative), and applying the FRR formula.

For FAR in the wakeword detection model, a negative dataset is necessary. This consists of audio samples that do not contain the wakeword, such as those from the tts-wwn category, where TTS-generated utterances do not include the keyword. By running the model on these negative samples, we can count the instances where the model falsely detects the wakeword (False Positive) and use those counts in the FAR formula.

When considering the voice authentication model, the positive dataset for FRR would consist of utterances from the authorized user (voice-authp), where the model should successfully verify the speaker's identity. FRR is then determined by how frequently the model fails (False Negative) to correctly authenticate the authorized user from this set. Conversely, for FAR in the voice authentication context, the negative dataset would include utterances from unauthorized users (voice-authn). The FAR measures how often the model incorrectly authenticates (False Positive) these unauthorized speakers as authorized.

To calculate the system accuracy, we will sweep the threshold from 0 to 1 in increments of 0.05 resolution, calculating the corresponding FRR and FAR values at each threshold. By plotting these FRR and FAR results on the same coordinate system, we will generate FRR and FAR curves. This approach not only enables us to visualize the trade-off between false rejections and false acceptances as the threshold changes but also provides insight into the system's performance under varying conditions. Ultimately, by identifying the point where the FRR and FAR curves intersect, we will determine the Equal Error Rate (EER), which serves as a key indicator of the system's accuracy and balance between security and usability.

# RESULTS

## 5.1 Wakeword Detection Model

We evaluated wakeword detection performance by sweeping threshold values from 0.0 to 1.0 for each model variant—voice_919, voice_919_vad, mix_919, and mix_919_vad—and measuring the False Reject Rate (FRR) and False Acceptance Rate (FAR). To determine the optimal threshold for each model, we identified the Equal Error Rate (EER) where FRR and FAR intersect. Table 5 summarizes the results.

**Table 5** Threshold Evaluation and Equal Error Rate (EER) of Wakeword Model

| Model | Optimal Threshold | EER (%) |
| --- | --- | --- |
| voice_919 | 0.10 | 22.61 |
| voice_919_vad | 0.05 | 22.95 |
| mix_919 | 0.05 | 20.52 |
| mix_919_vad | 0.05 | 16.79 |

As illustrated in Figure 3, voice_919 shows a high FRR at increasing thresholds, making it prone to rejecting genuine wakeword activations. It achieves an EER of 22.61% at a threshold of 0.10.

Figure 4 presents voice_919_vad, which applies voice activity detection (VAD) to filter out non-speech segments. While this reduces FAR, it also increases FRR, yielding an EER of 22.95% at a threshold of 0.05.

Figure 5 highlights mix_919, which benefits from incorporating TTS-based wakeword positives during training. This model achieves an EER of 20.52% at a threshold of 0.05, showing a more balanced trade-off between FRR and FAR compared to voice_919.

Figure 6 demonstrates mix_919_vad, which combines VAD with TTS-based wakeword training. It achieves the lowest EER of 16.79% at a threshold of 0.05, making it the most balanced model in the evaluation.

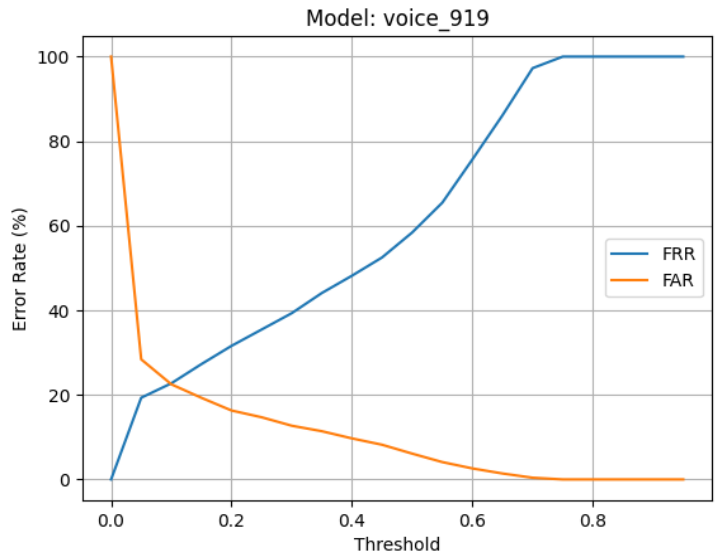

**Figure 3** FRR-FAR Curve of voice_919

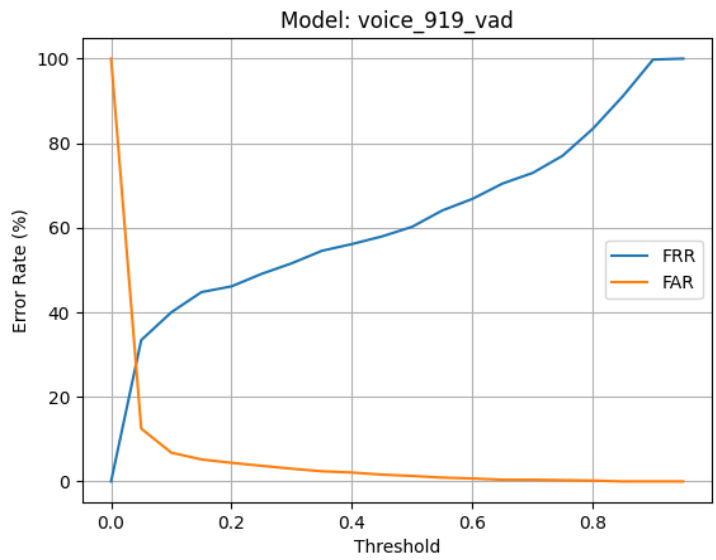

**Figure 4** FRR-FAR Curve of voice_919_vad

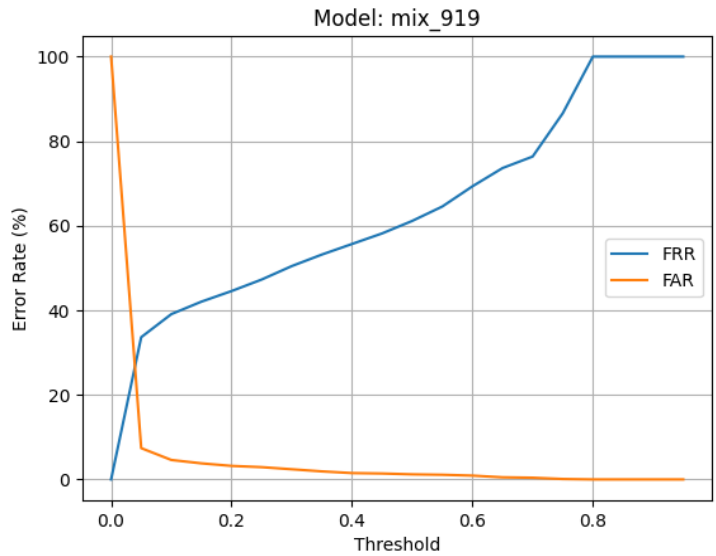

**Figure 5** FRR-FAR Curve of mix_919

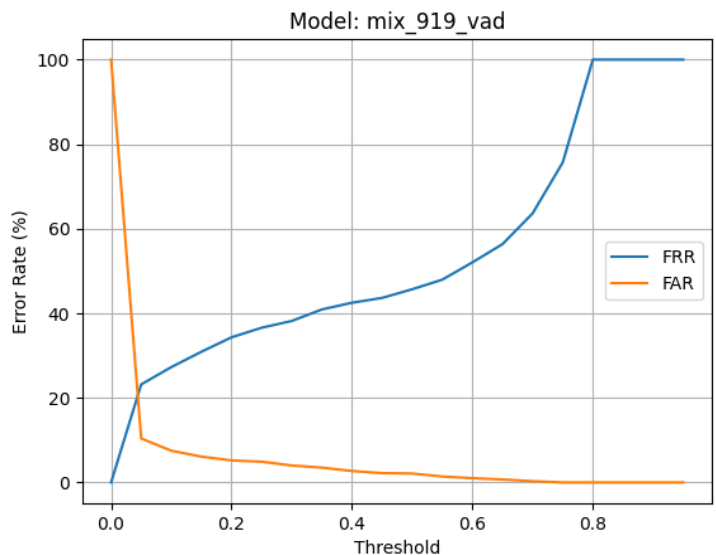

**Figure 6** FRR-FAR Curve of mix_919_vad

By comparing these optimal points, mix_919_vad at a threshold of 0.05 offers the best balance, achieving the lowest EER without significantly elevating FRR or FAR. This suggests that incorporating both VAD and TTS-based wakeword data improves wakeword detection robustness.

### 5.2 Speaker Authentication Model

We evaluated speaker authentication using two approaches: Approach A (96-dimensional embeddings) and Approach B (256-dimensional embeddings from Resemblyzer, x-vector, or ECAPA). The threshold sweep results for each model are shown in Figures 1-4, highlighting how the False Rejection Rate (FRR) and False Acceptance Rate (FAR) change across different thresholds. Table 6 describes the optimal threshold and Equal Error Rate (EER) of each models.

Table 6 Threshold Evaluation and Equal Error Rate (EER) of Speaker Auth Model

| Model | Optimal Threshold | EER (%) |
|---|---|---|
| Embedding_96d | 0.35 | 22.23 |
| Resemblyzer | 0.80 | 6.60 |
| x-vector | 0.95 | 10.37 |
| ECAPA | 0.40 | 9.54 |

As illustrated in Figure 7, the Embedding_96d model struggles to balance FRR and FAR, leading to a high EER of 22.23%. The performance is suboptimal due to its limited ability to capture speaker characteristics effectively.

Figure 8 highlights Resemblyzer, which achieves the lowest EER of 6.60% at a threshold of 0.80. The smooth transitions in FRR and FAR indicate that it provides a balanced performance, making it the most reliable among the evaluated models.

Figure 9 shows x-vector, which achieves an EER of 10.37% at a threshold of 0.95. While its FRR remains low over most threshold values, its FAR exhibits higher variance, making it less predictable in real-world scenarios.

Figure 10 presents ECAPA, which achieves an EER of 9.54% at a threshold of 0.40. This model demonstrates more stable FRR and FAR transitions compared to Embedding_96d, though its threshold tuning is more sensitive, requiring fine adjustments for practical deployment.

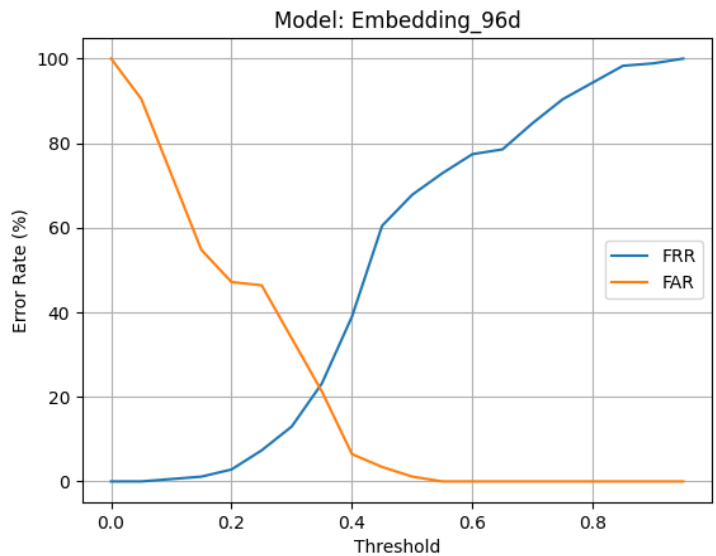

Figure 7 FRR-FAR Curve of Embedding_96d

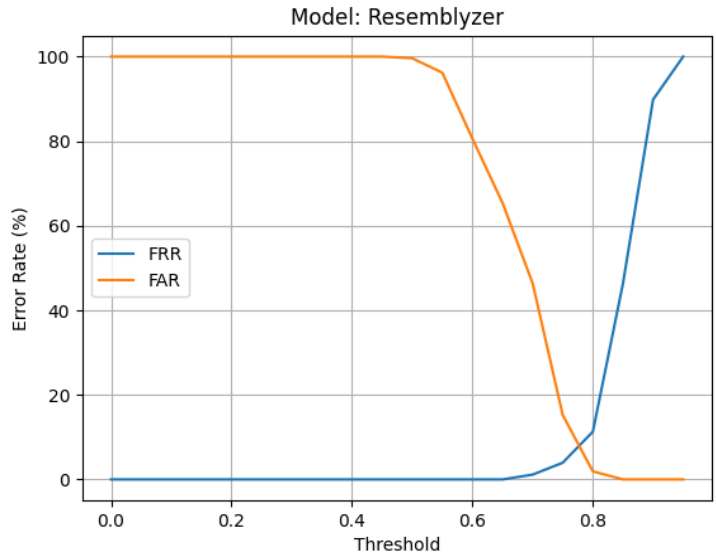

**Figure 8** FRR-FAR Curve of Resemblyzer

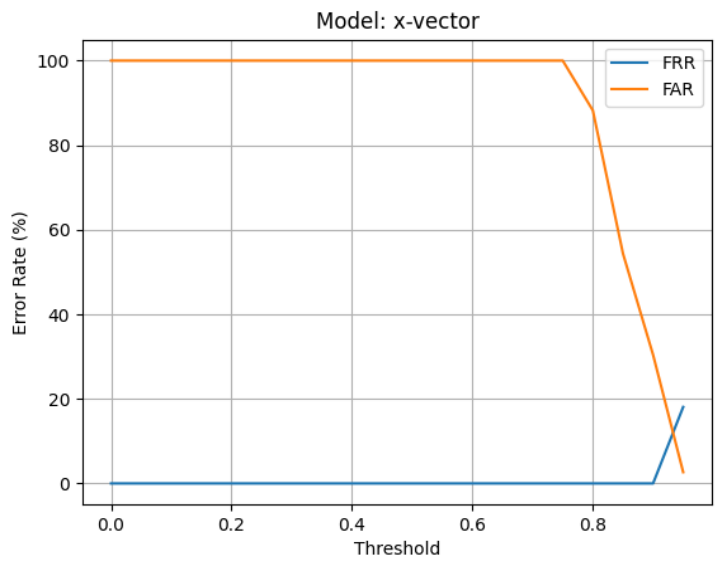

**Figure 9** FRR-FAR Curve of x-vector

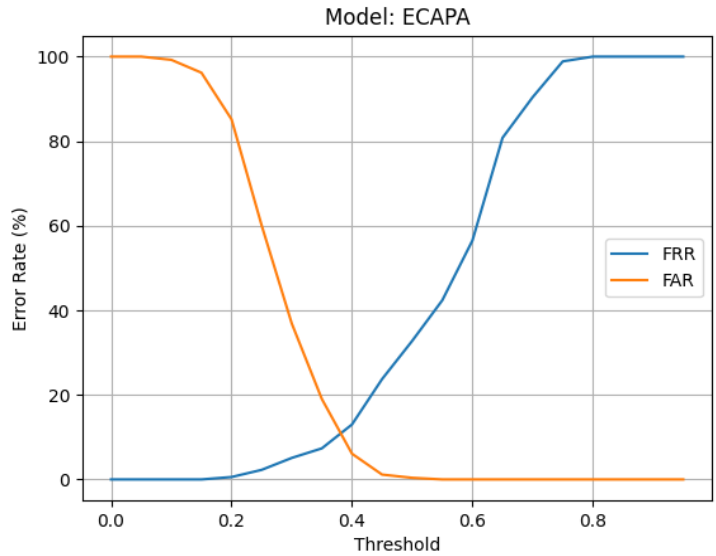

**Figure 10** FRR-FAR Curve of ECAPA

Comparing these optimal points, Resemblyzer at a threshold of 0.80 offers the best balance, achieving a low EER without significantly elevating either FRR or FAR. This confirms that Approach B (256D embeddings) is generally more effective for speaker authentication than Approach A (96D embeddings). For practical implementation, models like Resemblyzer and ECAPA are preferable due to their lower error rates and stable performance across varying thresholds.

## 5.3 CPU vs CUDA Acceleration

In addition to evaluating model accuracy and threshold optimization, we compared inference speed when running the speaker authentication module on the CPU versus leveraging CUDA acceleration. Table 7 summarizes the average loading time for the voice encoder model, as well as the mean authentication time per wakeword event across a series of test utterances (both successes and failures).

**Table 7** Performance Comparision of CPU and CUDA for Voice Authentication

| Configuration | Load Time | Avg. Auth Time | Notable Observations |
|---|---|---|---|
| CPU | 0.24 s | ~0.65 s | Quick Load, Slower Inference |
| GPU (CUDA) | 27.77 s | ~0.13 s | Longer Load, Fast Inference |

In summary, GPU acceleration significantly reduces per-wakeword authentication time, enhancing the responsiveness of AI assistants in continuous listening scenarios. However, developers must account for the longer initial model load on CUDA devices.

## DISCUSSION

The evaluation results underscore the inherent trade-offs in wakeword detection and speaker authentication, highlighting the Equal Error Rate (EER) as a crucial metric. For wakeword detection, the mix_919_vad model—combining Voice Activity Detection (VAD) with TTS-based training—demonstrated the most promising performance. It achieved a notably low EER of 16.79% at a threshold of 0.05, significantly outperforming models relying solely on speech data (e.g., voice_919 with 22.61% EER) or applying only VAD (voice_919_vad with 22.95% EER). By integrating both real and synthetic wakeword data along with effective silence trimming, mix_919_vad reduced both False Reject Rates (FRR) and False Acceptance Rates (FAR), leading to an improved balance between security and user convenience as evidenced by the lower EER.

In the domain of speaker authentication, high-dimensional (256D) embedding approaches delivered superior results. Among the evaluated models, Resemblyzer achieved the lowest EER of 6.60% at a threshold of 0.80, outperforming other alternatives such as x-vector (10.37% EER), ECAPA (9.54% EER), and the lower-dimensional 96D embedding model (22.23% EER). The low EER of Resemblyzer indicates that high-dimensional embeddings are more effective at capturing speaker characteristics, thereby simultaneously reducing FRR and FAR. This improvement is particularly significant for privacy, as a lower EER means a reduced likelihood of unauthorized voices being accepted.

Focusing on EER in our analysis suggests that dynamically adjusting thresholds—considering environmental factors like background noise or user-specific speaking patterns—can further refine both wakeword detection and speaker authentication accuracy. Small improvements in EER can have a significant impact on large user bases and over extended periods of system use. Dynamic threshold adjustments—guided by environmental factors such as background noise levels or user-specific patterns like individual speaking styles—can refine both wakeword detection and speaker authentication accuracy. Ongoing data collection from a variety of environments can further enhance robustness. Over time, the system could autonomously learn to fine-tune these thresholds, ensuring minimal inconvenience without sacrificing security.

If multiple consecutive authentication attempts fail, the system can implement various security measures to prevent unauthorized access. For example, after three unsuccessful authentications, the voice interface might require additional verification (e.g., a passcode) or temporarily block voice access to safeguard sensitive data. Additionally, it could send a notification to the administrator's phone, allowing immediate awareness of potential security risks. These measures not only enhance security but also ensure that the system remains robust in real-world deployments.

Overall, our end-to-end Korean wakeword and speaker authentication approach showcases promising performance but also illuminates areas—especially FRR improvements—for further innovation. By integrating a robust speaker verification stage, we ensure that simply uttering the wakeword "하나" does not automatically grant access to unauthorized individuals, thereby aligning with contemporary data privacy concerns.

## CONCLUSION

In this paper, we presented an end-to-end Korean wakeword detection and speaker authentication approach that addresses key privacy and performance challenges. By leveraging an FCN-based wakeword classifier, data augmentation strategies, and a CUDA-accelerated speaker verification pipeline, our system demonstrates the viability of Korean wakeword detection without sacrificing security. Experimental results highlight that incorporating TTS data and VAD can reduce false activations, and that 256-dimensional embeddings (e.g., Resemblyzer) outperform lower-dimensional ones for speaker authentication on resource-constrained hardware.

Moreover, our findings emphasize the delicate balance between FRR and FAR and underscore the importance of dynamic thresholding and robust data collection. While the proposed model already shows promising performance on resource-constrained hardware like the NVIDIA Jetson Nano, future work can explore advanced architectures, adaptive threshold adjustments, and on-device optimizations (pruning/quantization) to further enhance real-time performance and reduce EER. Ultimately, this work contributes a practical, privacy-centric framework for deploying AI voice assistants in Korean-language environments, offering insights that can be extended to other underrepresented languages as well.

# ACKNOWLEDGEMENT

We extend our sincere gratitude to No Hyun Woo (노현우), Park Jun Ha (박준하), Son Hae Su (손해수), Lee Da Hyeon (이다현), Lee Jae Hyeon (이재현), and Jo Jin Ah (조진아) for their generousity in providing the voice data used in this research. Their contributions were instrumental in enabling the development and evaluation of the Korean wakeword and speaker authentication systems presented in this paper.